\documentclass[%
reprint,
nofootinbib,
amsmath,amssymb,
aps,
prx,
]{revtex4-2}

\usepackage{graphicx}% Include figure files
\usepackage{dcolumn}% Align table columns on decimal point
\usepackage{bm}% bold math
\usepackage{subfigure}
\usepackage{floatrow}
\newfloatcommand{capbtabbox}{table}[][0.45\textwidth]
\usepackage{comment}
\usepackage{xcolor}
\usepackage{bbold}

\newcommand{\be}{\begin{equation}}
\newcommand{\ee}{\end{equation}}
\newcommand{\ba}{\begin{eqnarray}}
\newcommand{\ea}{\end{eqnarray}}

\newcommand{\snn}{\sigma_{nn}}
\newcommand{\snnb}{\boldsymbol{\sigma}}
\newcommand{\Sigmab}{\mathbf{\Sigma}}
\newcommand{\Rb}{\mathbf{R}}

\newcommand{\ave}[1]{\left\langle#1\right\rangle}
\newcommand{\mom}[1]{\langle#1\rangle}

\newcommand{\red}{}

\begin{document}
	
%\title[Article Title]{Article Title}
%\title{Extending intergranular normal-stress distributions using symmetries of isotropic linear-elastic polycrystalline materials}
\title{Extending intergranular normal-stress distributions using symmetries of linear-elastic polycrystalline materials}

\author{S.~El~Shawish}
\email{samir.elshawish@ijs.si}
\affiliation{Jo$\check{z}$ef Stefan Institute, SI-1000, Ljubljana, Slovenia}

\date{\today}

\begin{abstract}
Intergranular normal stresses (INS) are critical in the initiation and evolution of grain boundary damage in polycrystalline materials. To model the effects of such microstructural damage on a macroscopic scale, knowledge of INS is usually required statistically at each representative volume element subjected to various loading conditions. However, calculating INS distributions for different stress states can be cumbersome and time-consuming. This study proposes a new method to extend the existing INS distributions to arbitrary loading conditions using the symmetries of a polycrystalline material composed of randomly oriented linear-elastic grains with arbitrary lattice symmetry. The method relies on a fact that INS distributions can be accurately reproduced from the first (typically) ten statistical moments, which depend trivially on just {\red three stress} invariants and a few material invariants due to assumed isotropy and {\red material} linearity of the polycrystalline model. While these material invariants are complex averages, they can be extracted numerically from a few existing INS distributions and tabulated for later use. Practically, only {\red three} such INS distributions at properly selected loadings are required to provide all relevant material invariants for the first 11 statistical moments, which can then be used to reconstruct the INS distribution for arbitrary loading conditions. The proposed approach is demonstrated to be accurate and feasible for an arbitrarily selected linear-elastic material under various loading conditions.
\end{abstract}

\maketitle

% ==========================================================================
\section{Introduction}
\label{sec:intro}
% ==========================================================================

The ability to predict damage initiation and progression in structural materials heavily relies on understanding the local mechanical stresses present in the material. For certain aging processes, damage to the material begins at grain boundaries (GBs) where intergranular microcracks can form. Over time, these microcracks may grow along the GBs and merge into larger macroscopic cracks, ultimately threatening the structural integrity of the entire component under load. Therefore, it is essential to develop accurate models that can predict GB damage initiation based on sufficient knowledge of stresses present at GBs. An example of material aging mechanism that affects the GBs is an intergranular brittle fracture, which can occur in structural steels if the thermal history of a component leads to the segregation of phosphorus at the GBs \citep{thompson93}.

Various models have been developed to address damage initiation and propagation on different length and time scales. Among these, probabilistic modeling of GB damage initiation has emerged as a valuable tool for predicting and mitigating damage in components under specific loading conditions, and for identifying critical locations where damage is most likely to occur. This method involves quantifying the probability of GB damage initiation as a function of local (GB normal) stresses, material properties {\red and microstructure (grain morphology and crystallographic texture)}, and takes into account the variability and statistical distribution of these properties. One important input parameter for this type of modeling is the intergranular normal stress (INS) distribution, which needs to be estimated locally ({\textit i.e.}, on a scale of several hundreds of grains) for a given set of macroscopic loading conditions and material properties.

An example of a probabilistic modeling approach using INS distributions as an input parameter is the intergranular microstructure-informed brittle fracture (IG-MIBF) model, which is based on the original MIBF model \citep{forget2016} developed in the framework of the local approach to fracture defined by \citep{BEREMIN83} and used since then to predict the probability for brittle fracture of low alloy steels \citep{pineau2006}. The MIBF model is used to capture the effects of some microstructure characteristics such as the scatter of critical stresses induced by the carbide size distribution and the incompatibility stresses arising from the polycrystalline structure under applied deformation. Although the original MIBF model considers maximum principal stresses, the IG-MIBF model accounts for INS distributions, which typically require numerical evaluation at different loadings (defined by, \textit{e.g.}, stress triaxiality) using techniques such as crystal plasticity finite element (FE) simulations \citep{diard2002,diard2005,kanjarla,gonzalez2014,hure2016,elshawish2018} or crystal plasticity fast Fourier transform simulations \citep{lebensohn2012,disen2020,ren2022}. While these simulations can provide accurate results, they can become impractical and time-consuming when applied to numerous different loading cases. As a result, efforts have been made to identify the explicit role of external loading on INS distributions \citep{elshawish2021,elshawish2023}.

Despite recent developments, only a limited number of models exists in literature for predicting INS distributions in polycrystalline materials under general loading conditions. For instance, in \citep{elshawish2023} we proposed a simple semi-analytical model for elastic polycrystalline materials with arbitrary-shaped and randomly-oriented grains under uniform loading. This model provides algebraic expressions for local INS and corresponding uncertainties, which explicitly include the loading stress components. It was derived in a perturbative manner and systematically validated against FE simulations of various Voronoi microstructures with cubic {\red crystal lattices}. However, the model has not yet been calibrated for lower (non-cubic) lattice symmetries.

%The primary goal of this study is to develop a new alternative method for INS distribution modeling, based on the symmetry properties of a (quasi-)isotropic linear-elastic polycrystalline material composed of randomly oriented \textit{anisotropic} grains. The idea is to leverage these symmetries to establish an efficient and practical approach for extending the INS distributions from a limited number of loading conditions to any arbitrary uniform macroscopic stress state%
%
%\footnote{A non-uniform loading can be approximated as a quasi-uniform stress state within a sufficiently small material domain. The applicability of this study can be extended to such a domain when its volume exceeds that of a representative volume element (\textit{e.g.}, several thousand grains).}%
%
%. To achieve this, we derive a general expression for central statistical moments that explicitly incorporates the effect of loading stresses, enabling us to construct accurate INS distributions for various loading scenarios.

The primary goal of this study is to develop a new alternative method for INS distribution modeling, based on the {\red assumed properties of an \textit{isotropic} linear-elastic polycrystalline material. Here, the isotropy of a polycrystal is understood in statistical sense, emerging from a sufficiently large number of \textit{anisotropic} linear-elastic grains of arbitrary lattice symmetry, which are randomly oriented (providing zero crystallographic texture) and randomly shaped (providing zero morphological texture)%
\footnote{\red In principle, grain shapes and crystal lattice orientations can exhibit a correlation, provided that the isotropic response of the polycrystal is preserved.}%
.} The idea is to leverage these symmetries to establish an efficient and practical approach for extending the INS distributions from a limited number of loading conditions to any arbitrary uniform macroscopic stress state%
\footnote{A non-uniform loading can be approximated as a quasi-uniform stress state within a sufficiently small material domain. The applicability of this study can be extended to such a domain when its volume exceeds that of a representative volume element (\textit{e.g.}, several thousand grains).}%
. To achieve this, we derive a general expression for central statistical moments that explicitly incorporates the effect of loading stresses, enabling us to construct accurate INS distributions for various loading scenarios.

%The main goal of this study is to develop an alternative approach to INS distribution modeling, which would be based solely on symmetry arguments of a quasi-isotropic linear-elastic polycrystalline material. The idea is to use these symmetries to establish a viable and efficient method for extending accurately the INS distributions from few specific to arbitrary loading stresses. This is achieved by deriving a general expression for central statistical moments, with identifying an explicit role of the loading stresses, which are then used to construct the INS distributions at different loading.

The paper is structured as follows. We begin in Sec.~\ref{sec:method} by developing a general expression for the central moments of the INS distribution, and then present the strategy for extending the existing INS distributions to arbitrary loading stress. In Sec.~\ref{sec:val}, we demonstrate the feasibility of our method using the FE simulations. Finally, in Sec.~\ref{sec:conc}, we provide some concluding remarks. All technical details are given in the Appendices.
 
%The paper is structured as follows: in Sec.~\ref{sec:method} the proposed method is introduced, starting with the development of a general expression for the central moments of the INS distribution, and later continuing with the strategy to extend existing INS distributions from specific to arbitrary stress loadings. In Sec.~\ref{sec:val} the feasibility of the method is demonstrated using FE simulations. In Sec.~\ref{sec:conc} some concluding remarks are given. All technical details are deferred to the set of Appendices.

% ==========================================================================
\section{Method}
\label{sec:method}

% ==========================================================================
\subsection{Central moments}
\label{sec:cmom}

The purpose of this Section is to identify the role of the loading stress tensor $\Sigmab$ on the behavior of INS distribution and its respective statistical (central) moments in a general isotropic linear-elastic polycrystalline material subjected to $\Sigmab$. This will allow us later to develop a method to extend the existing INS distributions (calculated at specific $\Sigmab$s) to arbitrary loading conditions.

Since a polycrystalline aggregate under loading stress $\Sigmab$ is assumed to be isotropic%
\footnote{Although the grains are assumed anisotropic (with arbitrary {\red crystal} lattice symmetry), a polycrystalline aggregate becomes isotropic in the limit of a very large number of randomly {\red shaped grains and randomly oriented crystal lattices}.}%
, the INS distributions should be invariant to any aggregate rotation $\Rb$. Equivalently, the INS distributions should be invariant to any loading stress rotation $\Rb\Sigmab\Rb^T$. Selecting such $\Rb$ that $\Rb\Sigmab\Rb^T$ becomes diagonal, we may proceed with the most general loading stress
\begin{widetext}
\ba
\label{eq:load}
	\Sigmab&=&\left(
	\begin{array}{ccc}
		\Sigma_1 & 0 & 0\\
		0 & \Sigma_2 & 0\\
		0 & 0 & \Sigma_3
	\end{array}
	\right)=\Sigma_1
	\left(
	\begin{array}{ccc}
		1 & 0 & 0\\
		0 & 0 & 0\\
		0 & 0 & 0
	\end{array}
	\right)+
	\Sigma_2
	\left(
	\begin{array}{ccc}
		0 & 0 & 0\\
		0 & 1 & 0\\
		0 & 0 & 0
	\end{array}
	\right)+
	\Sigma_3
	\left(
	\begin{array}{ccc}
		0 & 0 & 0\\
		0 & 0 & 0\\
		0 & 0 & 1
	\end{array}
	\right)=
	\Sigma_1 \mathbf{I}_1+ \Sigma_2 \mathbf{I}_2 + \Sigma_3 \mathbf{I}_3
\ea
\end{widetext}
where $\Sigma_i$ are loading-stress eigenvalues and $\mathbf{I}_{i}$ unit-loading tensile stresses along the $i$ direction. If a local INS solution on an arbitrarily selected GB is $\snn^{(i)}$ for applied $\mathbf{I}_{i}$, then a local INS solution for applied $\Sigmab$ is
\be
\label{eq:lin}
	\snn=\Sigma_1 \snn^{(1)}+ \Sigma_2 \snn^{(2)}+ \Sigma_3 \snn^{(3)},
\ee
following the linearity principle (of the assumed generalized {\red(3D)} Hooke's law). Since all three directions $i$ are equivalent in an isotropic polycrystalline aggregate, the average INS scales with the trace of loading {\red (see also \citep{elshawish2023})},
\be
\label{eq:ave}
	\ave{\snn}=\sum_{i=1}^3 \Sigma_i \ave{\snn^{(i)}}=C \mathrm{tr}(\Sigmab)
\ee
where $\ave{\ldots}$ represents the averaging over all GBs {\red or a subset of GBs of specific type%
\footnote{\red We define a GB type by specifying crystallographic orientations of the two adjacent grains relative to the GB plane (see \citep{elshawish2021,elshawish2023} for more detail). It is also assumed that GBs of the same type are distributed randomly in the aggregate so that INS distribution obtained on them is invariant to arbitrary loading stress rotation ($\Sigmab\to\Rb\Sigmab\Rb^T$).}%
} in a polycrystalline aggregate and $C\equiv\ave{\snn^{(1)}}=\ave{\snn^{(2)}}=\ave{\snn^{(3)}}$ is a constant, {\red which depends on elastic material parameters and correlation strength between grain shapes and grain lattice orientations \citep{elshawish2023}. If the grains possess cubic lattice symmetry, then $C=1/3$ (see Appendix \ref{app0}).} Here we assumed that the grains are composed of the same material and they differ only by their crystallographic orientations.

The $m$-th central moment of the INS distribution is defined as
\be
\label{eq:cm}
	\mu^m=\ave{(\snn-\ave{\snn})^m}\big\vert_{\Sigmab}
\ee
for the applied stress $\Sigmab$. {\red It is convenient to decompose $\Sigmab$ into hydrostatic ($\Sigmab_{\text{hyd}}$) and deviatoric parts ($\Sigmab_{\text{dev}}$), $\Sigmab=\Sigmab_{\text{hyd}}+\Sigmab_{\text{dev}}$, and analyze both contributions to $\mu^m$ separately. Since $\Sigmab_{\text{hyd}}$ is fully symmetric, its effect on $\mu^m$ is significantly simplified. }

% ==========================================================================
%\subsubsection{\red Cubic lattice symmetry -- effect of $\Sigmab_{\text{dev}}$}
\subsubsection{\red Effect of $\Sigmab_{\text{dev}}$}
\label{sec:cubic}

%{\red The derivation of central moments simplifies considerably when the grains exhibit cubic lattice symmetry. Hence, cubic lattices are initially considered, with the results later extended to other (non-cubic) symmetries.}

%Since $\Sigmab_{\text{hyd}}$ is invariant to any orientation changes, $\Sigmab_{\text{hyd}}=\Rb\Sigmab_{\text{hyd}}\Rb^T$, its contribution to the local INS is constant {\red for each material point and thus for} each GB, $\snn=\Sigma_{\text{hyd}}$. {\red This implies that when grains exhibit cubic lattice symmetry $\Sigmab_{\text{hyd}}$ does not influence the width of the INS distribution, nor does it affect the corresponding central moments, $\mu^m=0\big\vert_{\Sigmab_{\text{hyd}}}$.} In this sense, the evaluation of the central moments can be reduced to

{\red We begin by deriving the central moments under the assumption of purely deviatoric loading. In fact, this loading is the sole contributor to $\mu^m$ when the grains posses cubic lattice symmetry%	
\footnote{\red In grains with cubic lattice symmetry, $\Sigmab_{\text{hyd}}$ provides a homogeneous stress field, $\snnb=\Sigmab_{\text{hyd}}$. This implies that $\Sigmab_{\text{hyd}}$ does not influence the width of the INS distribution, nor does it affect the corresponding central moments, $\mu^m=0\big\vert_{\Sigmab_{\text{hyd}}}$.}% 
. For non-cubic crystal lattices, however, the effect of hydrostatic loading is non-zero. This aspect is discussed later in Secs.~\ref{sec:noncubic1} and \ref{sec:noncubic2}.}%

The evaluation of the central moments can be reduced to

\be
\label{eq:rm}
	\mu_{\text{\red dev}}^m=\ave{(\snn-\ave{\snn})^m}\big\vert_{\Sigmab_{\text{dev}}}=\ave{\snn^m}\big\vert_{\Sigmab_{\text{dev}}}
\ee
when the applied stress $\Sigmab$ is replaced by the deviatoric stress $\Sigmab_{\text{dev}}$, which is traceless by definition, 
\begin{widetext}
\ba
\label{eq:dev}
	\Sigmab_{\text{dev}}&=&\Sigmab-\frac{1}{3}\mathrm{tr}(\Sigmab) \mathbb{1}_{3\times 3}\nonumber\\
	&=&\frac{2\Sigma_1-\Sigma_2-\Sigma_3}{3} \mathbf{I}_{1}+ \frac{-\Sigma_1+2\Sigma_2-\Sigma_3}{3} \mathbf{I}_{2}+ \frac{-\Sigma_1-\Sigma_2+2\Sigma_3}{3} \mathbf{I}_{3}
	\equiv\tilde{\Sigma}_1 \mathbf{I}_{1}+ \tilde{\Sigma}_2 \mathbf{I}_{2}+ \tilde{\Sigma}_3 \mathbf{I}_{3}.
\ea
\end{widetext}
%
%Since hydrostatic stress $\Sigmab_{\text{hyd}}$ has no influence on central moments {\red for cubic grain lattices}%
%
%\footnote{\red Note that Eq. (\ref{eq:rm}) applies also for non-cubic lattices when the applied loading is constrained to be purely deviatoric (or non-hydrostatic).}%
%
{\red In this respect}%
, only two {\red (out of three)} deviatoric stress invariants, $J_2\equiv\mathrm{tr}(\Sigmab_{\text{dev}}^2)/2$ and $J_3\equiv\mathrm{det}(\Sigmab_{\text{dev}})$, control the shape of the INS distribution. {\red In the case of cubic crystal lattices, any hydrostatic contribution enclosed in the} first stress invariant $I_1\equiv\mathrm{tr}(\Sigmab)= \mathrm{tr}(\Sigmab_{\text{hyd}})$, manifests only as a shift of the INS distribution%
\footnote{
	While only two deviatoric stress invariants $J_2$ and $J_3$ control the central moments, all three stress invariants, $I_1$, $I_2$ and $I_3$, participate in the expression for raw moments $\mom{\snn^m}\big\vert_{\Sigmab}$. For this reason, the central moments are preferred to raw moments.
}.

Using Eqs.~\eqref{eq:lin}, \eqref{eq:rm} and \eqref{eq:dev}, the $m$-th central moment can be expressed as
%
%\begin{widetext}
\ba
\label{eq:cm2}
	\mu_{\text{\red dev}}^m&=&\sum_{i=0}^m\sum_{j=0}^i \binom{m}{i}\binom{i}{j}
	\tilde{\Sigma}_1^j \tilde{\Sigma}_2^{i-j} \tilde{\Sigma}_3^{m-i}\nonumber\\
	&\times&\ave{\left(\snn^{(1)}\right)^j \left(\snn^{(2)}\right)^{i-j} \left(\snn^{(3)}\right)^{m-i}}.
\ea
%\end{widetext}
%
Applying again the statistical equivalence of three spatial directions, the material average term (\textit{i.e.}, independent of loading) in Eq.~\eqref{eq:cm2} becomes invariant to the permutation of the three indices 1,2 and 3,
%
%\begin{widetext}
\ba
	M(i,j,k)&\equiv&\ave{\left(\snn^{(1)}\right)^i \left(\snn^{(2)}\right)^j \left(\snn^{(3)}\right)^k}\nonumber\\
	&=&M(i,k,j)=M(j,i,k)=M(j,k,i)\nonumber\\
	&=&M(k,i,j)=M(k,j,i).
\ea
%\end{widetext}
%
Thus, a 3-parameter average $M(i,j,k)$ (with $i,j,k=0,1,2,\ldots$ and $i+j+k=m$) can be reduced to a 2-parameter average (using a commutative property of the multiplication of three indices),
\be
	M(i,j,k)\to M_m((i+1)(j+1)(k+1)).
\ee
This allows us to rewrite the $m$-th central moment in a symmetrized form
%
%\begin{widetext}
\ba
\label{eq:cm3}
	\mu_{\text{\red dev}}^m&=&\sum_{i=0}^m\sum_{j=0}^i \binom{m}{i}\binom{i}{j}
	\tilde{\Sigma}_1^j \tilde{\Sigma}_2^{i-j} \tilde{\Sigma}_3^{m-i}\nonumber\\ 
	&\times&M_m((j+1)(i-j+1)(m-i+1))
\ea
%\end{widetext}
%
where the averaging function $M_m$ depends solely on the (linear-elastic) properties of the grains {\red for the assumed zero crystallographic and zero morphological texture of the aggregate}. Moreover, if $\tilde{\Sigma}_1$, $\tilde{\Sigma}_2$ and $\tilde{\Sigma}_3$ in Eq.~\eqref{eq:cm3} are expressed in terms of the two (commonly chosen) deviatoric stress invariants $J_2$ and $J_3$, the expressions for $\mu_{\text{\red dev}}^m$ can be further simplified,
\ba
% using TM's I2 and I3 definition
%	&&\mu^1=0\to 0\\
%	&&\mu^2=I_2\left(M_2(3)-M_2(4)\right)\to I_2 M_{1,0}\nonumber\\
%	&&\mu^3=I_3\left(M_3(4)-3M_3(6)+2M_3(8)\right)\to I_3 M_{0,1}\nonumber\\ 
%	&&\mu^4=\frac{1}{2} I_2^2 \left(M_4(5)-4 M_4(8)+3 M_4(9)\right)\to I_2^2 M_{2,0}\nonumber\\
% using coomon I2 and I3 definition
	&&\mu_{\text{\red dev}}^1=0\to 0\\
	&&\mu_{\text{\red dev}}^2=2 J_2\left(M_2(3)-M_2(4)\right)\to J_2 M_{1,0}\nonumber\\
	&&\mu_{\text{\red dev}}^3=3 J_3\left(M_3(4)-3M_3(6)+2M_3(8)\right)\to J_3 M_{0,1}\nonumber\\ 
	&&\mu_{\text{\red dev}}^4=2 J_2^2 \left(M_4(5)-4 M_4(8)+3 M_4(9)\right)\to J_2^2 M_{2,0}\nonumber\\
	&&\quad\ \ \vdots\nonumber
\ea
%
\begin{comment}
\ba
\mu^1&=&0\nonumber\\
\mu^2&=&I_2\left(M_2(3)-M_2(4)\right)\nonumber\\
\mu^3&=&I_3\left(M_3(4)-3M_3(6)+2M_3(8)\right)\nonumber\\ 
\mu^4&=&\frac{1}{2} I_2^2 \left(M_4(5)-4 M_4(8)+3 M_4(9)\right)\nonumber\\
\mu^5&=&\frac{5}{6} I_2 I_3 \left(M_5(6)-5 M_5(10)+2 M_5(12)+8 M_5(16)-6 M_5(18)\right)\nonumber\\
\mu^6&=&\frac{1}{4} I_2^3 \left(M_6(7)-6 M_6(12)+15 M_6(15)-10 M_6(16)\right)\nonumber\\
&+&\frac{1}{3} I_3^2 \left(M_6(7)-6 M_6(12)-15 M_6(15)+20 M_6(16)+30 M_6(20)-60 M_6(24)+ 30 M_6(27)\right)\nonumber\\
\mu^7&=&\frac{7}{12} I_2^2 I_3 \left(M_7(8)-7 M_7(14)+9 M_7(18)-5 M_7(20)+12 M_7(24)-30 M_7(30)+20 M_7(32)\right)\nonumber\\
\mu^8&=&\frac{1}{8} I_2^4 \left(M_8(9)-8 M_8(16)+28 M_8(21)-56 M_8(24)+35 M_8(25)\right)\nonumber\\
&+&\frac{4}{9} I_2 I_3^2 \left(M_8(9)-8 M_8(16)-7 M_8(21)+49 M_8(24)-35 M_8(25)\right.\nonumber\\
&+&\left.35 M_8(28)-105 M_8(36)+35 M_8(40)+105 M_8(45)-70 M_8(48)\right)
\ea
\end{comment}
%
Here, a new parameter $M_{i,j}$ is introduced to collect all the material prefactors $M_m(k)$ in front of each $J_2^i J_3^j$ loading term (with $i,j=0,1,2,\ldots$ and $2i+3j=m$). In this way, the $\mu_{\text{\red dev}}^m$ finally simplifies to
\be
\label{eq:cm4}
	\mu_{\text{\red dev}}^m=\sum_{\substack{i,j\ge 0\\2i+3j=m}}J_2^i J_3^j M_{i,j}.
\ee
The first 20 central moments are presented in Tab.~\ref{tab:cm}.

\begin{table*}[t]
	\caption{\label{tab:cm}%
		The first 20 central moments $\mu_{\text{\red dev}}^m$ of the general INS distribution {\red for the assumed deviatoric stress loading $\Sigmab_{\text{dev}}$}, expressed in terms of the two deviatoric stress invariants $J_2\equiv\mathrm{tr}(\Sigmab_{\text{dev}}^2)/2$ and $J_3\equiv\mathrm{det}(\Sigmab_{\text{dev}})$, and material invariants $M_{i,j}$ with $2i+3j=m$.
	}
		\begin{tabular}{ll||ll}
			\hline
			m & $\mu_{\text{\red dev}}^m$ & m & $\mu_{\text{\red dev}}^m$\\
			%			\colrule
			\hline
			1 & 0 & 11 & $J_2^4 J_3 M_{4,1}+J_2 J_3^3 M_{1,3}$ \\
			2 & $J_2 M_{1,0}$ & 12 & $J_2^6 M_{6,0}+J_2^3 J_3^2 M_{3,2}+J_3^4 M_{0,4}$\\
			3 & $J_3 M_{0,1}$ & 13 & $J_2^5 J_3 M_{5,1}+J_2^2 J_3^3 M_{2,3}$\\
			4 & $J_2^2 M_{2,0}$ & 14 & $J_2^7 M_{7,0}+J_2^4 J_3^2 M_{4,2}+J_2 J_3^4 M_{1,4}$\\
			5 & $J_2 J_3 M_{1,1}$ & 15 & $J_2^6 J_3 M_{6,1}+J_2^3 J_3^3 M_{3,3}+J_3^5 M_{0,5}$\\
			6 & $J_2^3 M_{3,0}+J_3^2 M_{0,2}$ & 16 & $J_2^8 M_{8,0}+J_2^5 J_3^2 M_{5,2}+J_2^2 J_3^4 M_{2,4}$\\
			7 & $J_2^2 J_3 M_{2,1}$ & 17 & $J_2^7 J_3 M_{7,1}+J_2^4 J_3^3 M_{4,3}+J_2 J_3^5 M_{1,5}$\\
			8 & $J_2^4 M_{4,0}+J_2 J_3^2 M_{1,2}$ & 18 & $J_2^9 M_{9,0}+J_2^6 J_3^2 M_{6,2}+J_2^3 J_3^4 M_{3,4}+J_3^6 M_{0,6}$\\
			9 & $J_2^3 J_3 M_{3,1}+ J_3^3 M_{0,3}$ & 19 & $J_2^8 J_3 M_{8,1}+J_2^5 J_3^3 M_{5,3}+J_2^2 J_3^5 M_{2,5}$\\
			10 & $J_2^5 M_{5,0}+J_2^2 J_3^2 M_{2,2}$ & 20 & $J_2^{10} M_{10,0}+J_2^7 J_3^2 M_{7,2}+J_2^4 J_3^4 M_{4,4}+J_2 J_3^6 M_{1,6}$\\
			\hline
		\end{tabular}
\end{table*}
Finding a general expression for $\mu_{\text{\red dev}}^m$ {\red under purely deviatoric stress loading (note, however, that $\mu^m=\mu_{\text{\red dev}}^m$ for cubic crystal lattices)} by decoupling the loading ($J_2, J_3$) and material ($M$) contributions, is the first main result of this study. 

It is relevant to note that the simplicity of $\mu_{\text{\red dev}}^m$ in Eq.~\eqref{eq:cm4} results from the following assumed symmetries: (i) the isotropy of the polycrystalline material, (ii) the linearity of the employed Hooke's law and (iii) {\red the absence of hydrostatic loading (or its zero effect on the shape of INS distribution in the case of cubic lattice symmetry)}. Moreover, in the expression for $\mu_{\text{\red dev}}^m$, the two deviatoric stress invariants $J_2, J_3$ are independent parameters which are, however, bounded%
\footnote{To keep the stress components $\Sigma_{ij}$ real.}
by $J_2\ge 0$ and $|J_3|\le 2 J_2^{3/2}/\sqrt{27}$. Also, since odd central moments $\mu_{\text{\red dev}}^m$ ($m$ odd) scale with only odd powers of $J_3$, the sign inversion $J_3\to -J_3$ implies a reflection of the INS distribution about the $\ave{\snn}$ point, $\mathrm{PDF}(\snn-\ave{\snn})\to\mathrm{PDF}(-\snn+\ave{\snn})$, with $J_3=0$ denoting a symmetric INS distribution.

Having a relatively simple expression for $\mu_{\text{\red dev}}^m$ for a \textit{general} isotropic linear-elastic polycrystalline material subjected to {\red purely deviatoric} loading stress $\Sigmab_{\text{\red dev}}$ (\textit{i.e.}, $J_2$ and $J_3$), allows to extract the unknown material parameters $M_{i,j}$ from $N$ existing INS distributions obtained at few specific $\Sigmab_{\text{\red dev}}$. For example, if two ($N=2$) such INS distributions are available (\textit{e.g.}, from FE simulations or literature), along with their central moments, one can identify all $M_{i,j}$ of the first 11 (and also 13-th) central moments (see Tab.~\ref{tab:cm}), assuming that $J_2\ne 0$ and $J_3\ne 0$ of the $N=2$ existing INS distributions%
\footnote{Note that five lowest $M_{i,j}$, belonging to $\mu^m$ with $m=2,3,4,5$ and 7, can be obtained already from $N=1$ existing INS distribution. However, when more such INS distributions are available ($N>1$), one can calculate the five lowest $M_{i,j}$ from each distribution independently and average them to improve the accuracy of $M_{i,j}$.}.
With this, one can calculate the new central moments $\mu_{\text{\red dev}}^m$ for $m\le 11$ and arbitrary $\Sigmab_{\text{\red dev}}$ (\textit{i.e.}, $J_2$ and $J_3$). As demonstrated in Sec. \ref{sec:val}, $m\le 11$ is usually sufficient to accurately (re)construct the INS distributions.

% ==========================================================================
%\subsubsection{\red Non-cubic lattice symmetries -- effect of $\Sigmab_{\text{hyd}}$}
\subsubsection{\red Effect of $\Sigmab_{\text{hyd}}$}
\label{sec:noncubic1}

{\red Grains with lower (non-cubic) crystal lattice symmetries yield non-trivial local responses even under purely hydrostatic (symmetric) loading conditions, $\Sigmab=\Sigmab_{\text{hyd}}$. This indicates that the corresponding central moments are non-zero, $\mu^m\ne0\big\vert_{\Sigmab_{\text{hyd}}}$. Consequently, in the derivation of $\mu^m$, it is essential to consider a complete loading, $\Sigmab=\Sigmab_{\text{hyd}}+\Sigmab_{\text{dev}}$.
	
Following the same derivation steps as in Sec. \ref{sec:cubic}, the final expression for $\mu^m$, defined in Eq.~(\ref{eq:cm}) for arbitrary loading $\Sigmab$, simplifies to ($m>1$)
\be
\label{eq:cm5}
	\mu^m=\sum_{\substack{i,j,k\ge 0\\i+2j+3k=m}}I_1^i I_2^j I_3^k M_{i,j,k}
\ee
where all three stress invariants, $I_1\equiv\mathrm{tr}(\Sigmab)$, $I_2\equiv\left(\mathrm{tr}(\Sigmab)^2 -\mathrm{tr}(\Sigmab^2)\right)/2$ and $I_3\equiv\mathrm{det}(\Sigmab)$, multiply the unknown material prefactors $M_{i,j,k}$. Since several combinations ($i,j,k$) satisfy $i+2j+3k=m$ for a given $m$-th central moment, many more ($N\gg 1$) existing INS distributions are required to extract $M_{i,j,k}$ even for relatively low orders $m$,
%
%\be
%  \mu^3=I_1^3 M_{3,0,0} + I_1 J_2 M_{1,1,0} + J_3 M_{0,0,1},
%\ee
%
\ba
	&&\mu^1=0\\
	&&\mu^2=I_1^2 M_{2,0,0} + I_2 M_{0,1,0}\nonumber\\
	&&\mu^3=I_1^3 M_{3,0,0} + I_1 I_2 M_{1,1,0} + I_3 M_{0,0,1}\nonumber\\ 
	&&\mu^4=I_1^4 M_{4,0,0} + I_1^2 I_2 M_{2,1,0} + I_2^2 M_{0,2,0} + I_1 I_3 M_{1,0,1}\nonumber\\
	&&\quad\ \ \vdots\nonumber
\ea
For example, if four ($N=4$) existing INS distributions (obtained at different loadings $\Sigmab$ with $I_i\ne 0$) are available, it is possible to determine the corresponding $M_{i,j,k}$ for only the first four central moments. Clearly, the method outlined in Sec.~\ref{sec:cubic} for extracting $M_{i,j}$ from $\mu_{\text{\red dev}}^m$ proves impractical here for any realistic application.\\

To circumvent this, we propose an alternative path. The core idea is that on any given GB the two INS contributions%
%
%\footnote{\red While $\mathrm{tr}(\Sigmab_{\text{dev}})=0$, the corresponding $\mathrm{tr}(\snnb_{\text{dev}})\ne 0$ for non-cubic crystal lattices.}%
%
, $\snn=\snn^{\text{dev}}+\snn^{\text{hyd}}$, denoted as $\snn^{\text{dev}}$ for purely deviatoric loading $\Sigmab_{\text{dev}}$ and $\snn^{\text{hyd}}$ for purely hydrostatic loading $\Sigmab_{\text{hyd}}$, can be approximated as being independent of each other. This follows from the notion that $\snn^{\text{dev}}$ is strongly dependent on GB normal orientation $\mathbf{n}$, while $\snn^{\text{hyd}}$ is not (or much less) since there is no preferred direction in a (random) aggregate under purely symmetric (hydrostatic) loading%
\footnote{\red The assumed independence of $\snn^{\text{dev}}$ and $\snn^{\text{hyd}}$ can be easily tested in the FE aggregate by calculating a linear correlation between the two values on each GB using Pearson correlation coefficient $r$. For various $\Sigmab_{\text{dev}}$ and $\Sigmab_{\text{hyd}}$ used in Sec.~\ref{sec:valnoncubic}, we indeed found very low vales, $|r|<0.06$, which supports the assumption.}%
. Under this approximation, $\snn^{\text{hyd}}$ can be \textit{modeled} as stochastic variable with assumed normal probability distribution%
\footnote{\red The absence of a preferred direction under purely hydrostatic loading, coupled with the concept of a random crystallographic texture and random morphological texture, implies that $\mathrm{PDF}(\snn^{\text{hyd}})$ should be a symmetric function around its mean. From this perspective, a normal probability distribution appears to be an appropriate choice, as also supported by numerical results in Fig.~\ref{fig05}.
}
centered around the mean value for isotropic (or cubic crystal lattice) grains $I_1/3$, and with variance $\mu_{\text{hyd}}^2$,
\be
\label{eq:snnhyd}
	\snn^{\text{hyd}}\sim \mathcal{N}\left(\frac{I_1}{3}, \mu_{\text{hyd}}^2\right).
\ee
Since the response of grains is linear with respect to (hydrostatic) loading, the variance scales with $I_1^2$ so that $\mu_{\text{hyd}}^2=I_1^2 M_{2,0,0}$, where $M_{2,0,0}$ is unknown material parameter%
\footnote{\red Note that $M_{2,0,0}=0$ for grains with isotropic or cubic crystal lattices.}
to be extracted from a single existing INS distribution obtained at hydrostatic loading $\Sigmab_{\text{hyd}}=\frac{1}{3} I_1 \mathbb{1}_{3\times 3}$.
With this, all higher (even) central moments of normal distribution follow straightforwardly,
\be
\label{eq:cm6}
	\mu_{\text{hyd}}^m=\left\{\begin{array}{ll} 
		0 & ; m \text{ odd}\\
		I_1^m \left(M_{2,0,0}\right)^{m/2} (m-1)!! & ; m \text{ even}\end{array}\right..
\ee
%

% ==========================================================================
\subsubsection{\red Effect of $\Sigmab_{\text{hyd}}+\Sigmab_{\text{dev}}$}
\label{sec:noncubic2}

Since the (exact) approach outlined in Eq.~(\ref{eq:cm5}) is not feasible, the final (approximate) expression for the central moment $\mu^m$ is obtained from the probability theory, which states that the probability distribution of the sum of two or more independent random variables is the convolution of their individual distributions. In this respect,
\be
\label{eq:cm7}
	\mu^m\approx\sum_{i=0}^m \binom{m}{i} \mu_{\text{dev}}^i \mu_{\text{hyd}}^{m-i} 
\ee
where $\mu_{\text{dev}}^m$ is the (exact) $m$-th central moment of INS distribution evaluated for purely deviatoric loading, Eq.~(\ref{eq:cm4}), while $\mu_{\text{hyd}}^m$ is the (approximate) $m$-th central moment of INS distribution evaluated for purely hydrostatic loading, Eq.~(\ref{eq:cm6}). In the limit of vanishing hydrostatic contribution ({\textit i.e.}, when grains posses cubic lattice symmetry or $I_1=0$), the expression in Eq.~(\ref{eq:cm7}) correctly reduces to $\mu^m=\mu_{\text{dev}}^m$.\\

To summarize, we outline the following methodology for computing central moments for arbitrary crystal lattice symmetry and external loading $\Sigmab$, utilizing a limited number ($N+1$) of existing INS distributions evaluated at specific $\Sigmab$s:
\begin{enumerate}
	\item determine the unknown material parameters $M_{i,j}$ in $\mu_{\text{dev}}^m$ from $N$ existing INS distributions obtained at $N$ specific purely deviatoric loadings $\Sigmab_{\text{\red dev}}$ (with $I_1=0$),
	\item determine the unknown material parameter $M_{2,0,0}$ from a single existing INS distribution obtained at purely hydrostatic loading $\Sigmab_{\text{hyd}}=\frac{1}{3} I_1 \mathbb{1}_{3\times 3}$ (with $I_1\ne 0$),
	\item calculate the central moments $\mu^m$ from obtained prefactors $M_{i,j}$ and $M_{2,0,0}$ and arbitrary loading $\Sigmab$ (any $I_1$, $J_2$, $J_3$) using Eqs.~(\ref{eq:cm4}), (\ref{eq:cm6}), (\ref{eq:cm7}) and Tab.~\ref{tab:cm}.
\end{enumerate}

The proposed method is exact (limited by the accuracy of the inputs) for isotropic linear-elastic polycrystals composed of  grains that exhibit (i) cubic lattice symmetry under any external loading $\Sigmab$, or (ii) non-cubic lattice symmetry under purely deviatoric external loading $\Sigmab_{\text{dev}}$. In both cases, $\mu^m=\mu_{\text{dev}}^m$. As validated in Sec. \ref{sec:val}, the method performs very accurately also for most general crystal lattices and loadings.

}

\subsection{Construction of INS distribution}
\label{sec:constr}

The next step is to construct the INS distribution $\mathrm{PDF}(\snn)$ from first $K$ central moments $\mu^m$ ($m\le K$). There are various strategies how to get the spectra most representative for $K\to\infty$ (see, {\textit{e.g.}}, \citep{oitmaa84}), expecting a smooth function $\mathrm{PDF}(\snn)$. We follow here the procedure proposed by \citep{nickel74} to obtain the following expression (for details see Appendix \ref{app1}) 
\begin{widetext}
\ba
\label{eq:pdf}
	\mathrm{PDF}(\snn&-&\ave{\snn};K,P,\lambda)\approx
	-\frac{1}{\pi} \mathrm{Im}\left(\lim_{\epsilon\to 0}\left(\big[P/P\big]_{S(\xi;K,\lambda)}
%	\big\vert_{\xi\to\left(\omega-\sqrt{\omega^2-4\lambda^2}\right)/2-i\epsilon}\right)\right).
	\big\vert_{\xi\to\left((\snn-i\epsilon)\pm\sqrt{(\snn-i\epsilon)^2-4\lambda^2}\right)/2}\right)\right)
\ea
\end{widetext}
where $\big[P/P\big]_{S(\xi;K,\lambda)}$ is a diagonal Padé approximant (of order $P$, typically {$P\sim 6$}) to the modified moment series
\be
	S(\xi;K,\lambda)\equiv\sum_{m=0}^K \frac{G_m(\lambda)}{\xi^{m+1}}
\ee
about the point $\xi\to\infty$. The modified moments $G_m(\lambda)$, defined as
\be
	G_m(\lambda)\equiv\frac{1}{(m+1)!}\frac{\partial^{m+1}}{\partial{\xi^{m+1}}}\left(\frac{1}{\mu-1/\xi-\lambda^2\xi}\right)\Bigg\vert_0,
\ee
are linear combinations of central moments $\mu^i$ with prefactors $\lambda^{m-i}$ (with $0\le i\le m$) listed in Tab.~\ref{tab:G}%
\footnote{The expressions $\mom{\omega^i}$ in Tab.~\ref{tab:G} are indeed raw moments, which can be replaced by central moments $\mu^i$ to produce the INS distribution positioned at $\ave{\snn}=0$. However, to obtain the true INS distribution, an additional off-shift by $\ave{\snn}$ is required, see left-hand side of Eq.~\eqref{eq:pdf}.}.
In addition to the two newly introduced (integer) parameters $K$ and $P$ in Eq.~\eqref{eq:pdf}, the parameter $\lambda$ is a real number, which represents approximately the lower and upper bounds of the (centered) INS distribution as
\be
\label{eq:lamb}
	-2\lambda\le\snn-\ave{\snn}\le 2\lambda\quad\text{for all $\snn$.}
\ee
In practice, $\lambda$ should be as small as possible (while still obeying Eq.~\eqref{eq:lamb}), which is usually achieved by setting $\lambda$ to a fraction of von Mises loading stress $\Sigma_{\text{mis}}=\sqrt{3 J_2}$, \textit{e.g.}, {$\lambda/\Sigma_{\text{mis}}\sim 0.7$}. {This scaling ($\lambda\sim\Sigma_{\text{mis}}$) results from the fact that standard deviation of INS distribution scales as $\sqrt{\mu^2}\sim\sqrt{J_2}\sim\Sigma_{\text{mis}}$ (see Tab.~\ref{tab:cm}). However, the optimal values for $\lambda$ as well as $P$ depend (also) on the number of available central moments $K$. For a given number $N$ (and thus $K$) of available INS distributions, the fine-tuning of $\lambda$ and $P$ can be thus achieved by back-fitting of the $N$ reconstructed INS distributions obtained at specific loadings. The optimized $\lambda$ and $P$ can then be used to predict the INS distributions for other loadings.}

In Eq.~\eqref{eq:pdf}, the off-shift by $\ave{\snn}$ in the argument of $\mathrm{PDF}$ is required because the central moments are used in $G_m(\lambda)$ instead of raw moments (see Appendix \ref{app1}). Also, the "$\pm$" sign should be understood as "$+$" for $\snn\ge 0$ and "$-$" for $\snn<0$, while a reasonably small value of $\epsilon$ (\textit{e.g.}, $\epsilon/\Sigma_{\text{mis}}\sim 0.001$) should be taken instead of $\lim_{\epsilon\to 0}$ to provide smooth distributions in the range of interest, Eq.~\eqref{eq:lamb}.

\section{Validation of the method}
\label{sec:val}

The proposed method is validated against polycrystalline FE simulations of an artificial quasi-isotropic aggregate model (4000 randomly shaped and oriented grains) under various (uniform) loading conditions $\Sigmab$. More details about the FE simulations can be found in Appendix \ref{app2}. For demonstration purposes, 
%
%gamma-iron ($\gamma$-Fe) material properties with cubic lattice symmetry were used for the grains (see Table~\ref{tab:au}).
%
{\red two materials were selected for the grains: gamma iron ($\gamma$-Fe) with cubic crystal lattice symmetry (see Sec.~\ref{sec:valcubic}) and calcium sulfate (CaSO$_4$) with orthorhombic crystal lattice symmetry (see Sec.~\ref{sec:valnoncubic}). The corresponding material properties are listed in Tab.~\ref{tab:au}.}

\subsection{\red Cubic crystal lattice}
\label{sec:valcubic}

\begin{figure*}
	\centering
	\includegraphics[width=0.8\columnwidth]{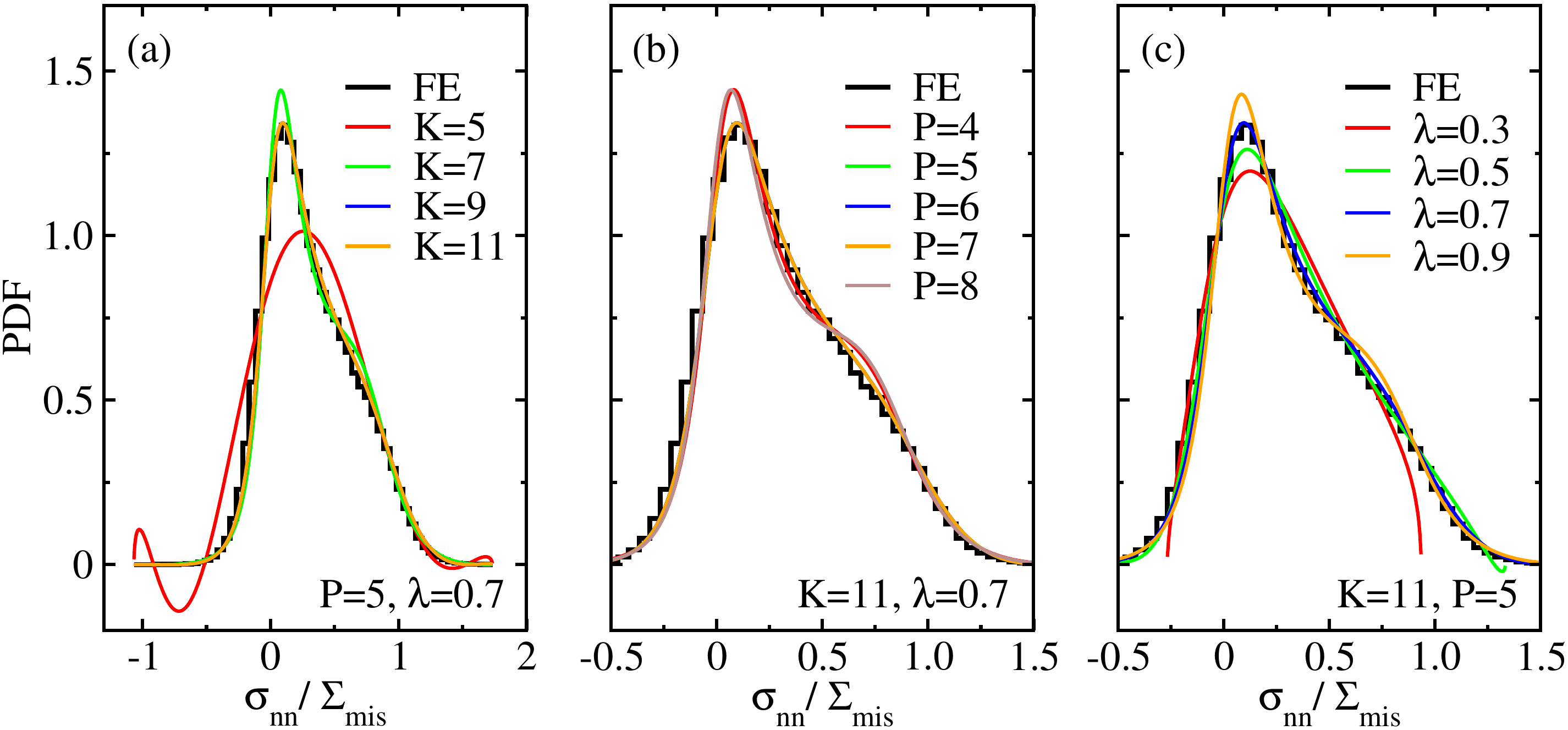}
	\caption{Feasibility of the method for the reconstruction of INS distribution $\mathrm{PDF}(\snn/\Sigma_{\text{mis}})$ from available first $K$ central moments. The original distribution (labeled FE) and its central moments $\mu^m$ ($m\le K$) were calculated from the FE simulation of $\gamma$-Fe polycrystal under tensile loading. The $\mu^m$ were then used to reconstruct the $\mathrm{PDF}$ using Eq.~\eqref{eq:pdf} with different parameter sets: (a) effect of $K$, (b) effect of $P$, (c) effect of $\lambda$ (in units of $\Sigma_{\text{mis}}$). All the lines were calculated with $\epsilon/\Sigma_{\text{mis}}=0.001$ in the range of $-2\lambda\le\snn-\ave{\snn}\le 2\lambda$.}
	\label{fig01}
\end{figure*}

Figure \ref{fig01} demonstrates the feasibility of the method presented in Sec.~\ref{sec:constr} for the reconstruction of $\mathrm{PDF}(\snn)$ from available first $K$ central moments $\mu^m {\red =\mu_{\text{dev}}^m}$ ($m\le K$). The moments $\mu^m$ were calculated numerically from the FE simulation of $\gamma$-Fe polycrystal under tensile loading ($\Sigmab=\mathbf{I}_1$) and then used to reconstruct the $\mathrm{PDF}$ by employing Eq.~\eqref{eq:pdf} with different parameter sets $(K, P, \lambda)$. The reconstructed $\mathrm{PDFs}$ show (Fig.~\ref{fig01}(a)) a rapid convergence with increasing $K$, reaching well-converged results already for $K\sim 9$. Also, the best performance of the method is obtained for $P\sim 6$ (\textit{e.g.}, $P=5,6$ or 7 provide the same result, see Fig.~\ref{fig01}(b)) and $\lambda/\Sigma_{\text{mis}}\sim 0.7$ (Fig.~\ref{fig01}(c)). The derived optimal values, $K\ge 9$, $P\sim 6$ and $\lambda/\Sigma_{\text{mis}}\sim 0.7$ should not change significantly when using other material properties%
\footnote{The most relevant material parameter here is (elastic) anisotropy of grains. More anisotropic materials produce wider INS distributions, thus larger $\lambda/\Sigma_{\text{mis}}$ is anticipated there.}.
\begin{figure*}
	\centering
	\includegraphics[width=0.8\columnwidth]{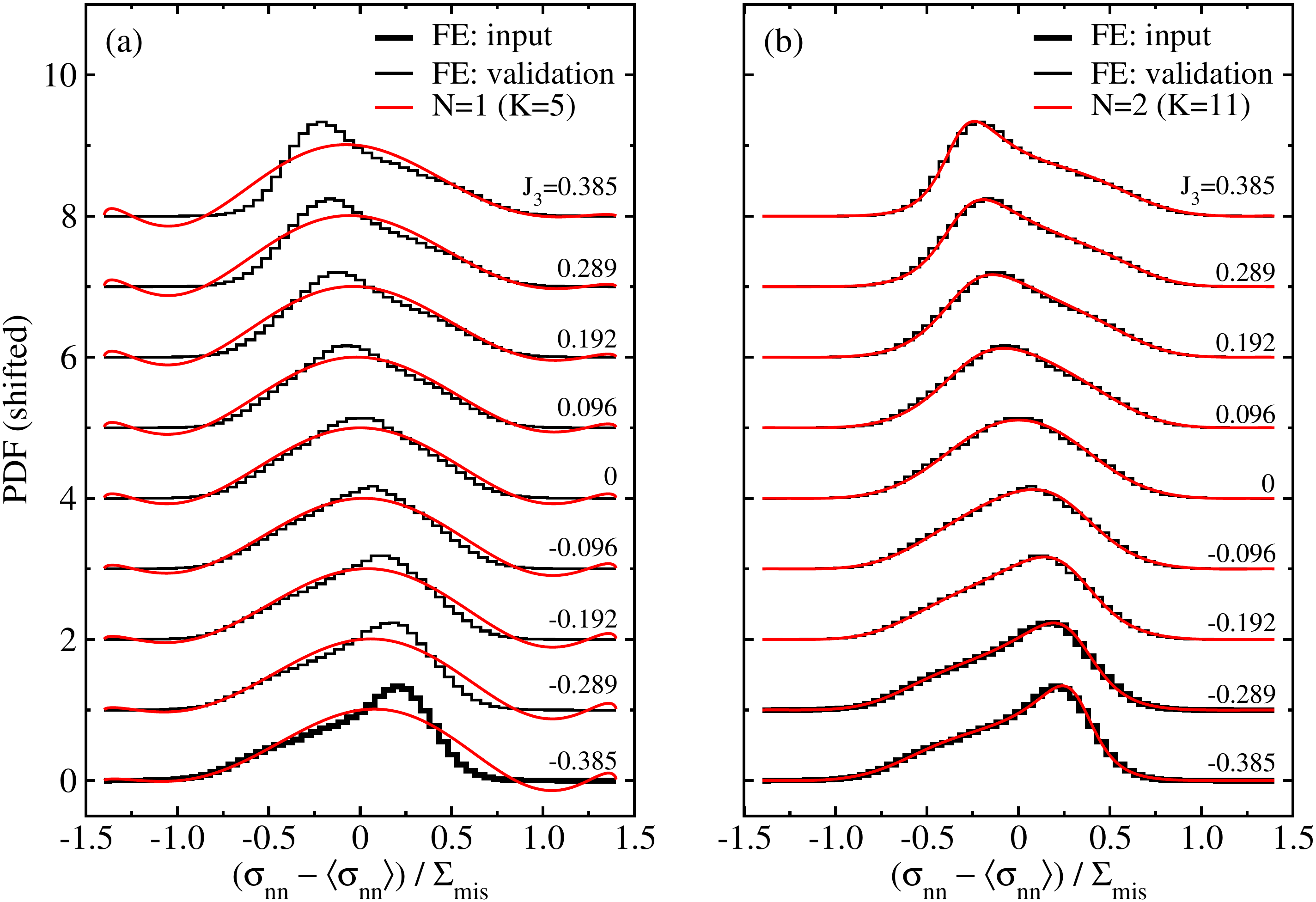}
	\caption{INS distributions (PDFs, shifted for clarity) for different loading conditions ($J_2=1$ and various $J_3$). A comparison is shown between PDFs obtained in FE simulations (labeled FE: validation) and PDFs predicted by the method (in red) using $N$ available distributions as an input (labeled FE: input): (a) $N=1$ and (b) $N=2$. The FE distributions were calculated from the FE simulations of $\gamma$-Fe polycrystal {\red (cubic crystal lattice)}. The predicted distributions were computed for: (a) $K=5$ and (b) $K=11$. In both cases, $P=5$, $\lambda/\Sigma_{\text{mis}}=0.7$, $\epsilon/\Sigma_{\text{mis}}=0.001$.}
	\label{fig02}
\end{figure*}

In the following, we present the feasibility of the method introduced in Sec.~\ref{sec:cmom} for the calculation of central moments (and corresponding PDFs) at arbitrary loadings $\Sigmab$ given the central moments at few ($N\sim 1$) specific $\Sigmab$s. Two cases are considered in Fig.~\ref{fig02}, one with $N=1$ and one with $N=2$ available input PDFs. Having only one such distribution at hand, restricts us to identifying $M_{i,j}$ of the first $K=5$ central moments (see Tab.~\ref{tab:cm}), which is obviously insufficient for accurate reconstruction and prediction (see Fig.~\ref{fig02}(a)). However, the situation is greatly improved when two PDFs become available (see Fig.~\ref{fig02}(b)). It this case, $K=11$ central moments are clearly sufficient to correctly reproduce and extend the existing two INS distributions to arbitrary loading conditions (defined by $J_2=1$ and varying $J_3$, see next paragraph). This holds true for any pair of input PDFs provided that their $|J_3|$s are different and nonzero. As practically identical results were obtained for $N=3$ (and $K=17$), they were omitted from Fig.~\ref{fig02} for brevity.
 
To exhaust different loading possibilities in a condensed manner, $\snn$ was off-shifted in Fig.~\ref{fig02} by $\ave{\snn}\sim I_1$ and normalized by von Mises stress $\Sigma_{\text{mis}}$. Since the latter scales with $J_2$, it is actually sufficient to fix $J_2$ to an arbitrary (positive) value and vary only $J_3$. In Fig.~\ref{fig02}, therefore, various possible PDF shapes (for a given material $\gamma$-Fe) were reconstructed by setting $J_2=1$ and varying $|J_3|\le 2/\sqrt{27}\sim 0.385$. 
This approach allows for the behavior of the $\mathrm{PDF}((\snn-\ave{\snn})/\Sigma_{\text{mis}})$ to be controlled by a single dimensionless parameter $J_3/J_2^{3/2}$, as opposed to the $\mathrm{PDF}(\snn)$, which requires three parameters ($I_1$, $J_2$ and $J_3$) to be specified.
%
%In this way, only one (dimensionless) parameter $J_3/J_2^{3/2}$ controls the behavior of $\mathrm{PDF}((\snn-\ave{\snn})/\Sigma_{\text{mis}})$, in comparison to three parameters $I_1, J_2, J_3$ controlling the $\mathrm{PDF}(\snn)$.

To summarize, the second key result of this study, illustrated in Fig.~\ref{fig02}, shows that {\red for cubic crystal lattices} we can accurately extend the INS distributions from specific to arbitrary loading conditions using only two input distributions evaluated at two specific loading conditions.

%Demonstrating in Fig.~\ref{fig02} that two input INS distributions evaluated at two specific loadings are sufficient in practice to extend them accurately to arbitrary loading conditions, is the second main result of this study. 

%How about going into plasticity? Eq. (2) does (probably) not hold anymore? Maybe $\Sigma_i\to\Sigma_i(\epsilon)$?

\subsection{\red Orthorhombic crystal lattice}
\label{sec:valnoncubic}

{\red

\begin{figure*}
	\centering
	\includegraphics[width=0.6\columnwidth]{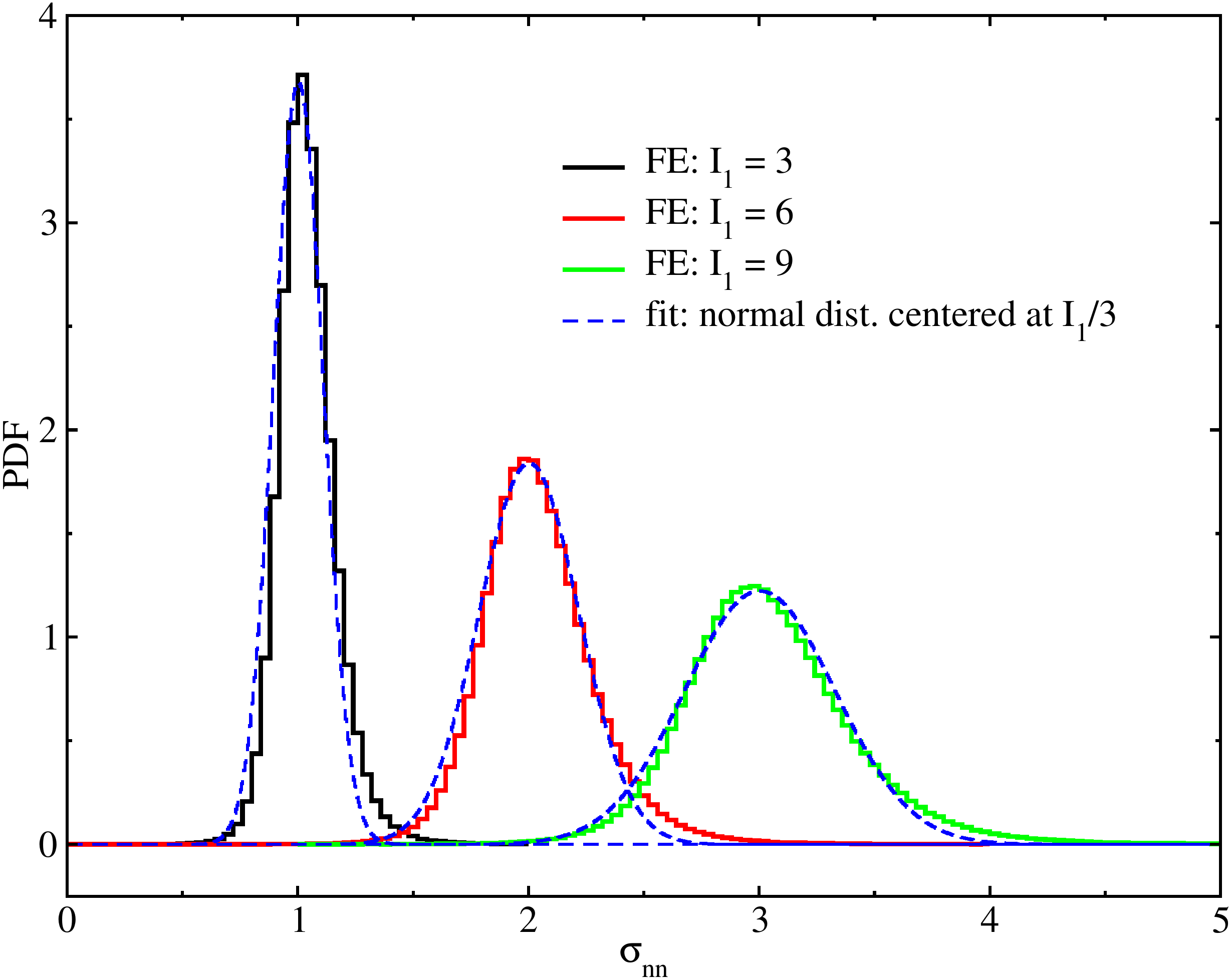}
	\caption{\red INS distributions for three purely hydrostatic loads ($I_1>0$, $J_2=J_3=0$), calculated from the FE simulations of CaSO$_4$ polycrystal {(orthorhombic crystal lattice)}. For comparison, normal distribution function is fitted to each INS distribution using single fitting parameter (variance).}
	\label{fig05}
\end{figure*}

In this section, the methodology for INS distribution reconstruction is validated for orthorhombic crystal lattice system CaSO$_4$ at various loading conditions. The emphasis is put on hydrostatic stress, which (contrary to cubic crystal lattices) provides non-zero contribution to central moments $\mu^m$. As shown in Fig.~\ref{fig05}, the effect of purely hydrostatic stress, $\Sigmab_{\text{hyd}}=\frac{1}{3} I_1 \mathbb{1}_{3\times 3}$, to INS distribution is indeed non-zero and can be accurately approximated%
\footnote{\red Small deviations observed at higher stresses are attributed to finite size effects, which emerge due to insufficient aggregate size (4000 grains) and/or too coarse FE mesh ($\sim$5 million elements).}
by a normal distribution centered at $I_1/3$ and with variance scaling as $I_1^2$. This supports the modeling assumption made in Sec.~\ref{sec:noncubic1}.

\begin{figure*}
	\centering
	\includegraphics[width=0.8\columnwidth]{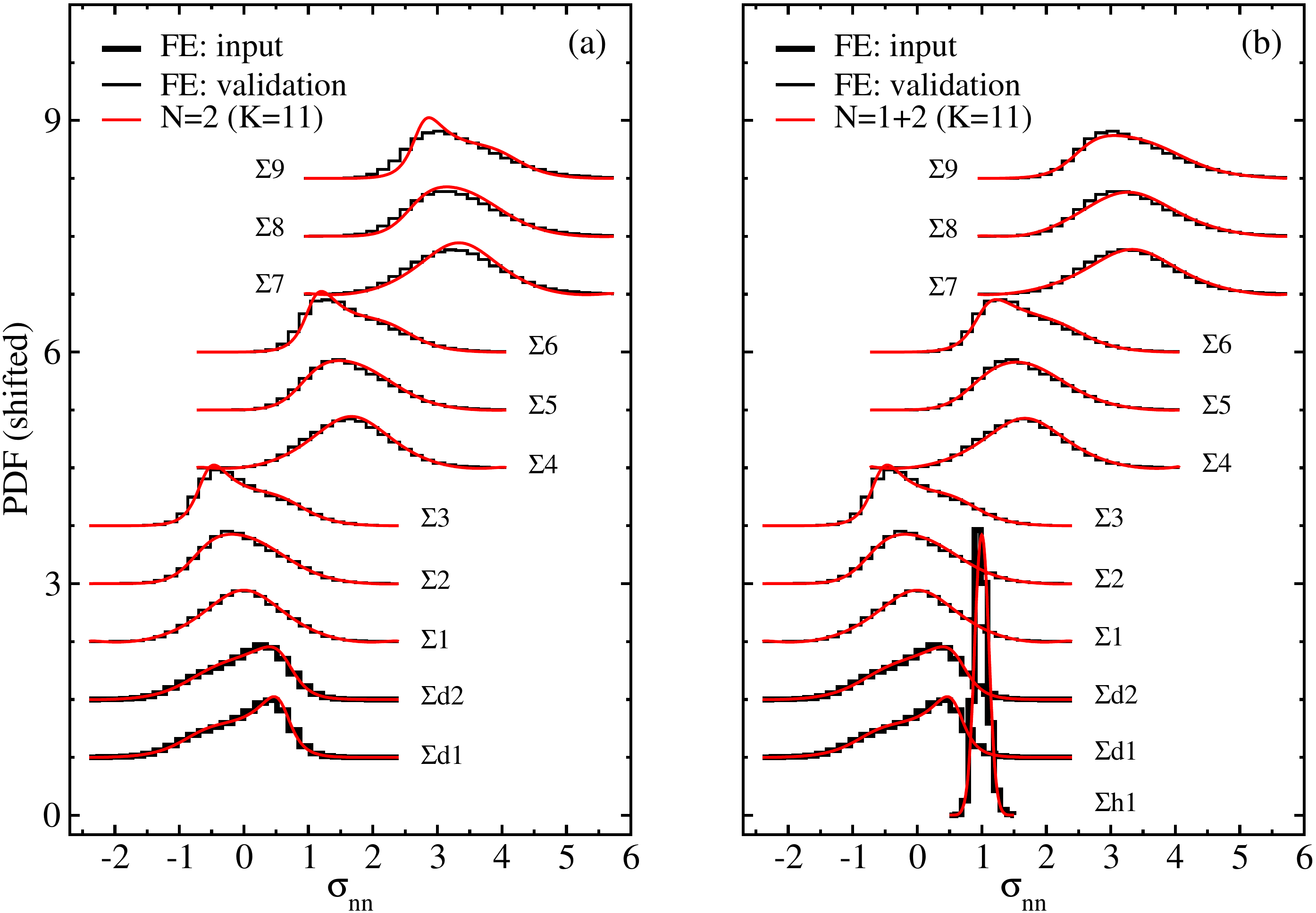}
	\caption{\red INS distributions (PDFs, shifted for clarity) for different loading conditions (see Tab.~\ref{tab:sigma}). A comparison is shown between PDFs obtained in FE simulations (labeled FE: validation) and PDFs predicted by the method (in red) using $N$ available distributions as an input (labeled FE: input): (a) $N=2$ (with two $\Sigmab_{\text{dev}}$ loadings) and (b) $N=1+2$ (with one $\Sigmab_{\text{hyd}}$ and two $\Sigmab_{\text{dev}}$ loadings). The FE distributions were calculated from the FE simulations of CaSO$_4$ polycrystal {(orthorhombic crystal lattice)}. The predicted distributions were computed for $K=11$, $P=5$, $\lambda=1.2$ (resp. 0.25 for $\Sigma$h1 loading), $\epsilon=0.001$.}
	\label{fig04}
\end{figure*}
\begin{table}[t]
	\caption{\label{tab:sigma}%
		\red Applied external stresses $\Sigma$n used in Fig.~\ref{fig04}. In the labeling of input-PDF stresses, one additional letter is used ("d" for purely deviatoric, and "h" for purely hydrostatic stress).
	}
	\begin{tabular}{llll||llll}
		\hline
		label & $I_1$ & $J_2$ & $J_3$ & label & $I_1$ & $J_2$ & $J_3$\\
		%			\colrule
		\hline
		$\Sigma$h1 & 1 & 0 & 0 & $\Sigma$4 & 5 & 1 & 0 \\
		$\Sigma$d1 & 0 & 1 & $-\frac{2}{3\sqrt{3}}$ & $\Sigma$5 & 5 & 1 & $\frac{1}{3\sqrt{3}}$ \\
		$\Sigma$d2 & 0 & 1 & $-\frac{1}{2\sqrt{3}}$ & $\Sigma$6 & 5 & 1 & $\frac{2}{3\sqrt{3}}$ \\
		$\Sigma$1 & 0 & 1 & 0 & $\Sigma$7 & 10 & 1 & 0 \\
		$\Sigma$2 & 0 & 1 & $\frac{1}{3\sqrt{3}}$ & $\Sigma$8 & 10 & 1 & $\frac{1}{3\sqrt{3}}$ \\
		$\Sigma$3 & 0 & 1 & $\frac{2}{3\sqrt{3}}$ & $\Sigma$9 & 10 & 1 & $\frac{2}{3\sqrt{3}}$ \\
		\hline
	\end{tabular}
\end{table}

The effect of general applied stress is analyzed in Fig.~\ref{fig04}, where two approaches are compared. While in Fig.~\ref{fig04}(a) two PDFs are considered as input ($N=2$), both evaluated at purely deviatoric stresses ($I_1=0$), in Fig.~\ref{fig04}(b) one additional input PDF ($N=1+2$) is taken at purely hydrostatic stress ($J_2=J_3=0$). In contrast to Fig.~\ref{fig02}, the resulting INS distributions are presented in absolute scale to facilitate the effect of $I_1$, $J_2$ and $J_3$ on PDF position and shape.

The results of Fig.~\ref{fig04} demonstrate that INS distributions can be accurately predicted for both deviatoric ($I_1=0$) and non-deviatoric applied stresses ($I_1\ne 0$). Best accuracy is achieved by employing the strategy outlined in Sec.~\ref{sec:noncubic2} for the estimation of central moments under both $\Sigmab_{\text{hyd}}$ and $\Sigmab_{\text{dev}}$ (see Fig.~\ref{fig04}(b)). However, when the effect of $\Sigmab_{\text{hyd}}$ is neglected (\textit{i.e.}, by omitting the $\Sigma$h1 input), the agreement between FE simulations and predictions may reduce, especially at larger values $|I_1|$ (see top curves in Fig.~\ref{fig04}(a)). This validates the proposed methodology also for non-cubic crystal lattices.

Similar to the $\gamma$-Fe case shown in Fig.~\ref{fig02}, also for CaSO$_4$ the $K=11$ central moments are sufficient to correctly reproduce and extend the existing INS distributions to arbitrary loading conditions. As practically identical results were obtained for $N=1+3$ (and $K=17$), they were omitted from Fig.~\ref{fig04} for brevity.

In summary, for isotropic linear-elastic polycrystals with non-cubic crystal lattices three input INS distributions evaluated at three specific loading conditions are sufficient to accurately extend the INS distributions to arbitrary loads.

}

% ==========================================================================
\subsection{\red Limitations on the applicability of the method}
\label{sec:app}
% ==========================================================================

{\red 
The limitations of the proposed method, such as the assumption of linear elasticity of the grains and the absence of crystallographic and morphological texture in the aggregate, restrict its applicability to studying INS (distributions) in untextured aggregates under predominantly elastic loads. This could be particularly relevant for predicting the \textit{onset of GB damage} in various materials. Many failure mechanisms, which result in GB crack initiation, often begin in the elastic regime of the grains. Examples include intergranular stress corrosion cracking, hydrogen embrittlement, radiation-induced embrittlement, low-temperature brittle fracture, and intergranular corrosion. In all these cases, damage (and consequently cracking) can occur at GBs without significant plastic deformation. By examining GB normal stresses in this regime, it is possible to predict the onset of these failures, which is crucial for ensuring material reliability and safety.

Although the method requires both crystallographic and morphological textures to be absent for an isotropic polycrystal response, it does not limit the potential correlation between grain shape and the orientation of the crystal lattice. This allows the method to be used for studying INS (distributions) not only on random GBs but also on special GBs, where the two neighboring grains have fixed crystal lattice orientations relative to the GB plane. Such local arrangements of lattice orientations typically provide a certain strength to a GB, which can be compared to the corresponding INS for damage modeling. While many interesting results have been obtained on cubic crystal lattices \citep{elshawish2021,elshawish2023}, the responses on such GBs in less symmetric crystals are less known. In this regard, the proposed method could prove to be useful. 
}

% ==========================================================================
\section{Conclusion}
\label{sec:conc}
% ==========================================================================

In this study, we have derived a general expression for the $m$-th central moment of the intergranular-normal-stress (INS) distribution when evaluated in an {\red isotropic} linear-elastic polycrystalline material under uniform {\red applied stress}. These central moments depend on {\red three stress} invariants and only a few (generally unknown) material invariants, due to {\red material linearity and rotational} symmetry of a polycrystal. Based on these findings, we have developed a new strategy to extend existing INS distributions from specific stress loadings to arbitrary loading conditions. The method relies on the fact that INS distributions can be accurately (re)constructed using only few first central moments (typically ten) whose material invariants can be extracted numerically from the existing INS distributions. Our study has demonstrated the accuracy and feasibility of this approach for two selected linear-elastic materials {\red with different crystal lattice symmetries}, showing that {\red in a general case} only {\red three} input INS distributions evaluated at {\red three} specific {\red applied stresses} are sufficient for accurately extending them to arbitrary loading conditions.
These results have important implications for probabilistic modeling of grain boundary damage initiation on a component scale, particularly in the presence of complex stress states that can arise in front of propagating macroscopic discontinuities such as cracks.

% ==========================================================================
\section*{Acknowledgments}
% ==========================================================================
The author gratefully acknowledges financial support provided by Slovenian Research Agency (grant P2-0026). {\red The author also thanks the anonymous reviewer for recognizing and highlighting the limitation of the original manuscript.}

%\newpage

% ==========================================================================
\appendix
% ==========================================================================

% ==========================================================================
\section{\red Average INS for cubic crystal lattices}
\label{app0}

{\red

For statistically isotropic polycrystal with linear material response, the local INS and its average follow  Eqs.~(\ref{eq:load}), (\ref{eq:lin}) and (\ref{eq:ave}). 

If the grains of the polycrystal exhibit also cubic lattice symmetry, a homogeneous solution is obtained for the local stresses under purely hydrostatic external loading, $\snnb=\Sigmab_{\text{hyd}}$ and, consequently,
\be
	\snn=\frac{1}{3}\mathrm{tr}(\Sigmab_{\text{hyd}}).
\ee
Since hydrostatic loading corresponds to $\Sigma_1=\Sigma_2=\Sigma_3$ in Eq.~(\ref{eq:lin}), it follows that
\be
\label{eq:aq1}
	\snn^{(1)} + \snn^{(2)} + \snn^{(3)}=1.
\ee
This relation holds locally on every GB. From this and Eq.~(\ref{eq:ave}), it follows immediately that
\be
\label{eq:aq2}
	\ave{\snn}=\frac{1}{3} \mathrm{tr}(\Sigmab).
\ee

It is important to highlight that this derivation does not rely on any assumptions regarding the relationship between the grain shape and the orientation of the crystal lattice. The only prerequisites are: statistical isotropy of a polycrystal, linear material response and cubic crystal lattice symmetry.

Moreover, according to numerical simulations (not shown), the above result seems to remain valid also for lower crystal lattice symmetries, provided that averaging $\ave{\ldots}$ is performed over random GBs%
\footnote{\red We have been unable to provide a mathematical proof for this claim. However, one approach would be to replace the right hand side of Eq.~(\ref{eq:aq1}) with $1+f$, where $f$ is a random variable with symmetric PDF around zero. This seems a feasible assumption since there is no preferred direction in random aggregate under purely symmetric (hydrostatic) load (see also Sec.~\ref{sec:noncubic1}). Assuming this, Eq.~(\ref{eq:aq2}) follows naturally also for any other crystal lattice symmetry.}%
.

}

% ==========================================================================
\section{Details on the construction of INS distribution from raw moments}
\label{app1}
% ==========================================================================
We follow a procedure of Ref.~\citep{oitmaa84} developed for the frequency-dependent relaxation function $F(\omega)$ of the transverse Ising model in three dimensions. Function $F(\omega)$ stands here for our off-shifted%
\footnote{Due to using central moments instead of raw moments.}
INS distribution $\mathrm{PDF}(\snn-\ave{\snn})$ that we seek. 

Starting with the Laplace transform of the time-dependent relaxation function $F(t)$ with $\mathrm{Re}(s)>0$,
\be
\label{eq:a1}
	\hat{F}(s)=\int\limits_0^\infty F(t) e^{-s t} dt
\ee
and using a Fourier transform
\be
\label{eq:a2}
	F(t)=\frac{1}{\sqrt{2\pi}}\int\limits_{-\infty}^\infty F(\omega') e^{i \omega' t} d\omega',
\ee
one can integrate over $t$ for the assumed $s=i z$ with $\mathrm{Im}(z)<0$ to obtain
\be
\label{eq:a3}
	-i \sqrt{2\pi} \hat{F}(i z)\equiv \tilde{F}(z)=\int\limits_{-\infty}^\infty \frac{F(\omega')}{\omega'-z} d\omega'
\ee
where $\tilde{F}(z)$ is an analytic function of a complex variable $z$ in the lower half-plane. Applying series expansion of the above integrand about $z\to\infty$ and integrating over $d\omega'$, we get
\be
\tilde{F}(z)=-\sum_{m=0}^\infty \int\limits_{-\infty}^\infty \frac{F(\omega')(\omega')^m}{z^{m+1}} d\omega'=-\sum_{m=0}^\infty \frac{\mom{\omega^m}}{z^{m+1}}
\ee
where $\mom{\omega^m}$ is the $m$-th (raw) moment of $F(\omega)$. However, if we assume in Eq.~\eqref{eq:a3} that $z=\omega - i \epsilon$ (with both $\omega$ and $\epsilon>0$ real), then as $\epsilon\to 0$ we obtain (using Sokhotski–Plemelj theorem)
\ba
\label{eq:a4}
	\tilde{F}(\omega)&=&\lim_{\epsilon\to 0} \tilde{F}(z)\nonumber\\
	&=&\int\limits_{-\infty}^\infty \frac{F(\omega')}{\omega'-\omega} d\omega' -i\pi F(\omega)\nonumber\\
	&=&-\sum_{m=0}^\infty \frac{\mom{\omega^m}}{\omega^{m+1}}-i\pi F(\omega).
\ea
Thus 
\ba
\label{eq:a6}
	F(\omega)&=&-\frac{1}{\pi} \mathrm{Im}\left(\tilde{F}(\omega)\right)\nonumber\\
	&=&-\frac{1}{\pi} \mathrm{Im}\left(\lim_{\epsilon\to 0} \tilde{F}(z)\right)\nonumber\\
	&=&\frac{1}{\pi} \mathrm{Im}\left(\lim_{\epsilon\to 0}\sum_{m=0}^\infty \frac{\mom{\omega^m}}{(\omega-i\epsilon)^{m+1}}\right).
\ea
A direct approximation of the moment series in Eq.~\eqref{eq:a6} by, for example, Padé approximants is doomed to failure as one obtains only a set of poles which represents the branch cut along the $\mathrm{Re}(z)$ axis. The imaginary piece of $\tilde{F}(\omega)$ is simply a set of $\delta$ functions rather than a continuous function which we wish to obtain.

In order to circumvent this difficulty, \citep{nickel74} first carried out a nonlinear transformation
\be
\label{eq:a7}
%	\omega\to \xi+\lambda^2/\xi
	z\to \xi+\lambda^2/\xi
\ee
where $2\lambda$ is the length of the branch cut on the $\mathrm{Re}(z)$ axis which is here assumed to be centered at $\omega=0$. In the complex $\xi$ plane the branch cut is mapped onto a circle of radius $\lambda$. The interior of the circle contains the unphysical sheet of $\tilde{F}(\omega)$ to which any spurious singularities are confined. The exterior of the circle is the physical sheet. 

Using a nonlinear transformation of Eq.~\eqref{eq:a7} and applying series expansion of the integrand in Eq.~\eqref{eq:a3} about $\xi\to\infty$ and integrating over $d\omega'$, we get
\ba
\label{eq:a8}
	F(\omega)&=&-\frac{1}{\pi} \mathrm{Im}\left(\lim_{\epsilon\to 0} \tilde{F}(z)\right)\nonumber\\
%	=-\frac{1}{\pi} \mathrm{Im}\left(\lim_{\epsilon\to 0}\sum_{m=0}^\infty \frac{G_m(\lambda)}{(\xi-i\epsilon)^{m+1}}\right),
	&=&-\frac{1}{\pi} \mathrm{Im}\left(\lim_{\epsilon\to 0}\sum_{m=0}^\infty \frac{G_m(\lambda)}{\xi^{m+1}}\right),
\ea
where a modified-moment function $G_m(\lambda)$ is introduced as a linear combination of the raw moments $\mom{\omega^i}$ up to order $m$ ($0\le i\le m$), 
\begin{widetext}
\ba
\label{eq:a8b}
	G_m(\lambda)&=&\frac{1}{(m+1)!}\int\limits_{-\infty}^\infty \frac{\partial^{m+1}}{\partial{\xi^{m+1}}}\left(\frac{1}{w'-1/\xi-\lambda^2\xi}\right)\Bigg\vert_0 F(\omega') d\omega'\nonumber\\
	&=&\frac{1}{(m+1)!}\ave{\frac{\partial^{m+1}}{\partial{\xi^{m+1}}}\left(\frac{1}{w-1/\xi-\lambda^2\xi}\right)\Bigg\vert_0},
\ea
\end{widetext}
see Tab.~\ref{tab:G}.

\begin{table*}[t]
	\caption{\label{tab:G}%
		The first 10 expressions for the modified-moment function $G_m(\lambda)$ defined in Eq.~\eqref{eq:a8b}.
	}
	\begin{tabular}{ll}
		\hline
		m & $G_m(\lambda)$ \\
		%			\colrule
		\hline
		0 & $-1$ \\
		1 & $-\mom{\omega}$ \\
		2 & $\lambda^2-\mom{\omega^2}$ \\
		3 & $2\lambda^2\mom{\omega}-\mom{\omega^3}$ \\
		4 & $-\lambda^4+3\lambda^2\mom{\omega^2}-\mom{\omega^4}$ \\
		5 & $-3\lambda^4 \mom{\omega}+4\lambda^2 \mom{\omega^3}-\mom{\omega^5}$ \\
		6 & $\lambda^6-6\lambda^4 \mom{\omega^2}+5\lambda^2 \mom{\omega^4}-\mom{\omega^6}$ \\
		7 & $4\lambda^6 \mom{\omega}-10\lambda^4 \mom{\omega^3}+6\lambda^2 \mom{\omega^5}-\mom{\omega^7}$ \\
		8 & $-\lambda^8+10\lambda^6 \mom{\omega^2}-15\lambda^4 \mom{\omega^4}+7\lambda^2 \mom{\omega^6}-\mom{\omega^8}$ \\
		9 & $-5\lambda^8 \mom{\omega}+20\lambda^6 \mom{\omega^3}-21\lambda^4 \mom{\omega^5}+8\lambda^2 \mom{\omega^7}-\mom{\omega^9}$ \\
		10 & $\lambda^{10}-15\lambda^8 \mom{\omega^2}+35\lambda^6 \mom{\omega^4}-28\lambda^4 \mom{\omega^6}+9\lambda^2 \mom{\omega^8}-\mom{\omega^{10}}$ \\
		\hline
	\end{tabular}
\end{table*}
From the inverse of Eq.~\eqref{eq:a7} we obtain (using "$+$" for $\omega\ge0$ and "$-$" for $\omega<0$)
\begin{widetext}
\be
\label{eq:a9}
	F(\omega)=-\frac{1}{\pi} \mathrm{Im}\left(\lim_{\epsilon\to 0}\sum_{m=0}^\infty
%	\frac{G_m(\lambda)}{\left(\left(\omega-\sqrt{\omega^2-4\lambda^2}\right)/2-i\epsilon\right)^{m+1}}\right).
	\frac{G_m(\lambda)}{\left(\left((\omega-i\epsilon)\pm\sqrt{(\omega-i\epsilon)^2-4\lambda^2}\right)/2\right)^{m+1}}\right).
\ee
\end{widetext}
Although the radius of convergence of the above series is increased compared to that of Eq.~\eqref{eq:a6}, one can further improve the behavior of $F(\omega)$ for finite number ($m\le K$) of available moments $\mom{\omega^m}$ in $G_m(\lambda)$ by employing the Padé approximants to the series in Eq.~\eqref{eq:a8}, which do not have poles on the branch cut in the range $-2\lambda<\omega<2\lambda$. Using a diagonal Padé approximant $\big[P/P\big]_{S(\xi;K,\lambda)}$ to series $S(\xi;K,\lambda)$ about the point $\xi\to\infty$, we finally obtain
\begin{widetext}
\ba
\label{eq:a10}
	S(\xi;K,\lambda)&\equiv&\sum_{m=0}^K \frac{G_m(\lambda)}{\xi^{m+1}}\\
	F(\omega;K,P,\lambda)&\approx&-\frac{1}{\pi} \mathrm{Im}\left(\lim_{\epsilon\to 0}\left(\big[P/P\big]_{S(\xi;K,\lambda)}
%	\big\vert_{\xi\to\left(\omega-\sqrt{\omega^2-4\lambda^2}\right)/2-i\epsilon}\right)\right).
	\big\vert_{\xi\to\left((\omega-i\epsilon)\pm\sqrt{(\omega-i\epsilon)^2-4\lambda^2}\right)/2}\right)\right).
\ea
\end{widetext}
%
%In practice, the best results for $F$ are obtained for {$P\sim 6$} and reasonably small $\epsilon/\lambda\sim 0.001$.
%
In practice, the  $\lim_{\epsilon\to 0}$ is replaced by a sufficiently small $\epsilon$ (\textit{e.g.}, $\epsilon/\lambda\sim 0.001$) to provide a smooth $F(\omega)$ in the range $-2\lambda<\omega<2\lambda$.

% ==========================================================================
\section{Finite element aggregate model}
% ==========================================================================
\label{app2}

\begin{figure}
	\centering
	\includegraphics[width=0.9\columnwidth]{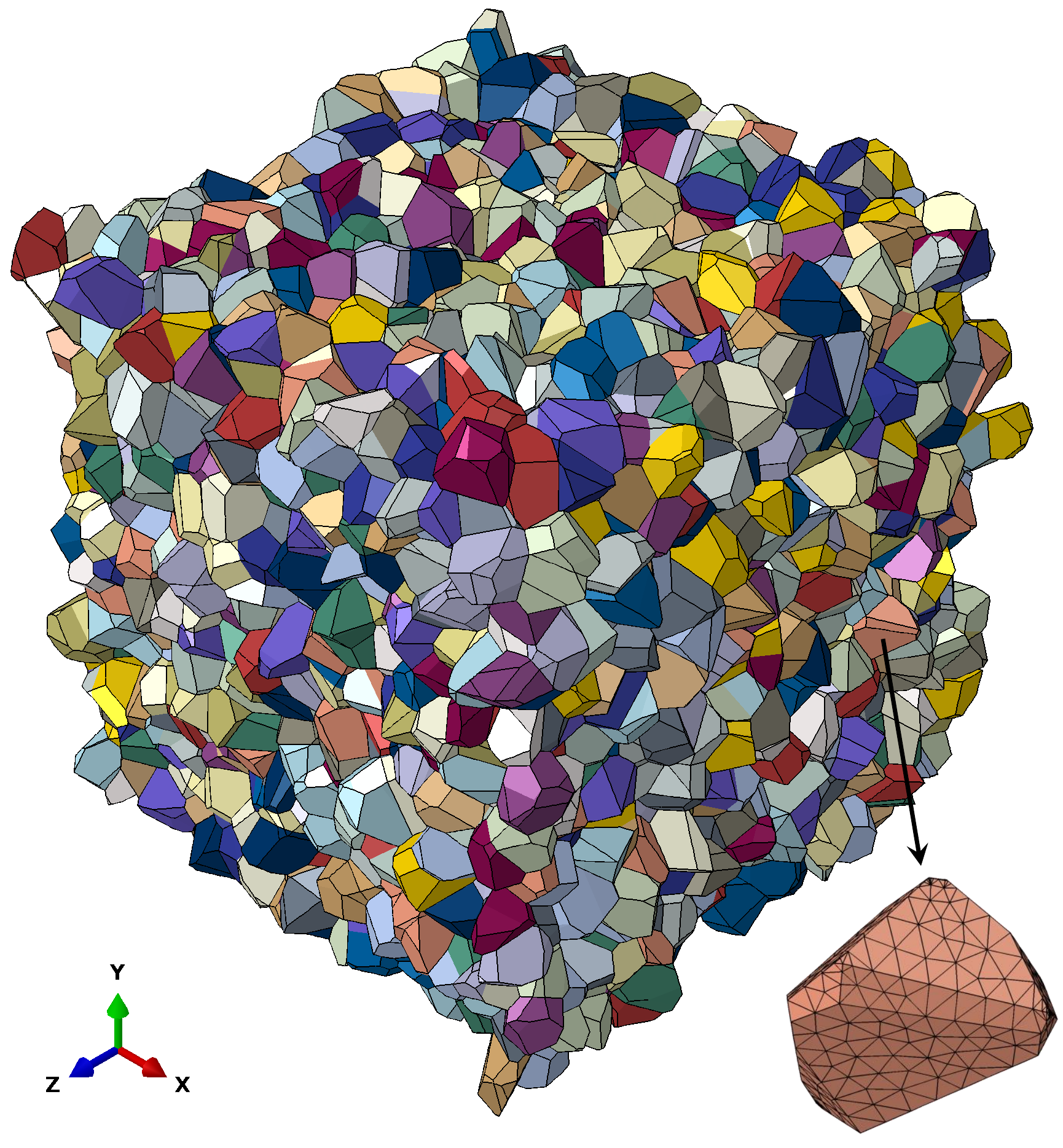}
	\caption{Periodic Voronoi aggregate model with 4000 grains and 31154 GBs. Different grains are denoted by different colors. FE mesh is shown for one selected grain.}
	\label{fig03}
\end{figure}

A polycrystalline aggregate model with 4000 randomly shaped and randomly oriented grains was generated upon Voronoi tessellation~\citep{neper} with periodic microstructure in all three spatial directions \citep{elshawish2019}.
A FE mesh was generated with quadratic tetrahedral elements ($\sim$5 million total elements of type C3D10) to preserve the geometry of the grains. The aggregate model used in this study is shown in Fig.~\ref{fig03}. A general uniform loading $\Sigmab$ was applied to the aggregate such that $\Sigma_{ij}=\mom{\sigma_{ij}}$ for the averages taken over the entire volume of the aggregate model. Since grains were assumed ideally elastic, a unit loading was applied using a small strain approximation.

The constitutive equations of the generalized Hooke's law were solved with FE solver Abaqus~\citep{abaqus}. Numerically calculated stress field $\sigma_{ij}$ was then used to obtain one $\snn$ value at each GB-element facet by first averaging the six Cauchy stresses $\sigma_{ij}$ at nearby Gauss points within the two elements sharing the facet and then projecting the averaged stress $\bar{\sigma}_{ij}$ onto the facet normal $n_i$, $\snn=\sum_{ij}\bar{\sigma}_{ij} n_i n_j$. It was verified that the aggregate size and FE mesh density were sufficiently large to produce quasi-isotropic responses and negligible finite size effects in the calculation of INS distributions%
\footnote{\red To test the isotropy of the FE aggregate model, three different effective stiffness values ($E_1$, $E_2$, $E_3$) were calculated for two materials by applying tensile loading along three orthogonal directions. For $\gamma$-Fe: $E_1=187.5$ GPa, $E_2=190.5$ GPa, $E_3=190.8$ GPa. For CaSO$_4$: $E_1=77.2$ GPa, $E_2=76.4$ GPa, $E_3=76.0$ GPa. In both cases, the maximum relative difference is smaller than 2\%.}%
.

The grains were assumed to be made {\red either} of gamma iron ($\gamma$-Fe) {\red or calcium sulfate (CaSO$_4$)} with material properties listed in Tab.~\ref{tab:au}. This choice was made arbitrarily for the purpose of demonstrating the feasibility of the method in Sec.~\ref{sec:val}.

\begin{table*}[h]
	\caption{\label{tab:au}
		Elastic constants $C_{ij}$ (in GPa, Voigt notation) of $\gamma$-Fe \citep{bower} {\red and CaSO$_4$} \citep{simmonswang} single crystals.}
		\begin{tabular}{lllllllllll}
			\hline
			Crystal & {\red symmetry} & $C_{11}$ & $C_{12}$ & $C_{13}$ & $C_{22}$ & $C_{23}$ & $C_{33}$ & $C_{44}$ & $C_{55}$ & $C_{66}$\\
			%			\colrule
			\hline
			$\gamma$-Fe  & {\red cubic} &197.5    & 125.0  & & & & & 122.0 & &\\
			{\red CaSO$_4$}  & {\red orthorhombic} & 93.82   & 16.50  & 15.20 & 185.45 & 31.73 & 111.80 & 32.47 & 26.53 & 9.26\\
			\hline
		\end{tabular}
\end{table*}

\bibliography{spebib2}

\end{document}